\documentclass[aps,prc,twocolumn,showpacs,showkeys,amsmath,amssymb]{revtex4}
\usepackage{graphicx,dcolumn,bm}

\newcommand{\kf}{k_{\rm F}}
\newcommand{\as}{a_s}
\newcommand{\re}{r_e}
\newcommand{\be}{\begin{equation}}
\newcommand{\ee}{\end{equation}}
\newcommand{\fm}{\, \text{fm}}

\newcommand{\mev}{\, \text{MeV}}
\newcommand{\gcq}{\, \text{g}/\text{cm}^{3}}
\newcommand{\vlk}{V_{{\rm low}\,k}}

\begin{document}

\title{The Virial Equation of State of Low-Density Neutron Matter}
\author{C.J. Horowitz}
\email[E-mail:~]{horowit@indiana.edu}
\author{A. Schwenk}
\email[E-mail:~]{schwenk@indiana.edu}
\affiliation{Nuclear Theory Center and Department of Physics, 
Indiana University, Bloomington, IN 47408}

\begin{abstract}
We present a model-independent description of low-density neutron matter
based on the virial expansion. The virial equation of state provides a
benchmark for all nuclear equations of state at densities
and temperatures where the interparticle separation is large compared 
to the thermal wavelength. We calculate the second virial coefficient 
directly from the nucleon-nucleon scattering phase shifts. Our results 
for the pressure, energy, entropy and the free energy correctly include
the physics of the large neutron-neutron scattering length. We find that, as 
in the universal regime, thermodynamic properties of neutron matter 
scale over a wide range of temperatures, but with a significantly reduced 
interaction coefficient compared to the unitary limit.
\end{abstract}

\pacs{21.65.+f, 26.50.+x, 97.60.Bw, 05.70.Ce}
\keywords{Low-density neutron matter, equation of state, virial expansion}

\maketitle

\section{Introduction}

Matter composed of neutrons and protons exhibits remarkable properties.
Cold neutron matter is close to a scaling regime for densities $n 
\lesssim 1/10 \, n_0$, where $n_0 = 0.16 \, \fm^{-3}$ denotes nuclear 
saturation density.
In this regime, thermodynamic properties of neutron matter scale only with 
the density, and for the energy per particle $E/N$ one has
approximately~\cite{Vijay,dEFT}
\be
\frac{E}{N} \approx 0.5 \, \biggl( \frac{E}{N} \biggr)_{\rm free} = 0.5 \,
\frac{3 (3 \pi^2)^{2/3}}{10 m} \, n^{2/3} \,,
\label{coldneut}
\ee
where $m$ is the nucleon mass.
In contrast, below saturation density, the energy per particle of symmetric 
nuclear matter is independent of the density $E/A \approx -16 \mev$, and 
nucleons cluster into nuclei and larger structures. The physics of nuclear 
matter is therefore a crossover from a gas of nucleons to homogeneous matter,
where nuclei and larger clusters coexist with the nucleon gas over a wide 
range of intermediate densities.
Nuclear clustering is due to the competition of nuclear binding, entropy and
the Coulomb repulsion.

With this rich interplay, it is very important to develop a reliable 
formalism for nucleonic matter over densities and temperatures 
that are experimentally or observationally accessible. In this 
paper, we present a model-independent description of neutron matter 
based on the virial expansion. We have previously applied the
virial expansion to low-density nuclear matter composed of neutrons, 
protons and alpha particles~\cite{vEOSnuc}.
The virial equation of state presents a controlled application of
nucleon scattering data to densities and temperatures where the 
interparticle separation is large compared to the thermal wavelength. 
For neutron matter, the relevant densities are
$n \lesssim 4 \cdot 10^{11} \, (T/\text{MeV})^{3/2} \gcq$,
where $T$ is the temperature. The virial equation of state
therefore provides important constraints on the physics of the
neutrinosphere in supernovae, where one has $T \approx 4 \mev$ 
and $n \sim 10^{11} \gcq \sim 1/1000 \, n_0$~\cite{sn1987a1,sn1987a2}.

Properties of dilute Fermi gases can also be studied in laboratory 
experiments with trapped atoms. Cold atom experiments provide
exciting access to neutron matter in the universal low-density
regime, where the scattering length $\as$ is large compared to
the interparticle separation $\kf \as \gg 1$ and the effective
range is small $\kf \re \ll 1$. Under these conditions, there 
are no length scales associated with the interaction, and the 
only dimensionful scale is the Fermi momentum $\kf$. Therefore, the
system will exhibit universal behavior, where all macroscopic 
observables are given by powers of $\kf$
multiplied by universal factors. For example, under these conditions 
the energy per particle $E/N$ for cold gases of atomic $^6$Li, $^{40}$K 
or neutrons with equal populations of two spin states is
\be
\frac{E}{N} = \xi \biggl( \frac{E}{N} \biggr)_{\rm free} = \xi \,
\frac{3 \kf^2}{10 m} \,,
\label{unieos}
\ee
where the universal factor $\xi$ is a number. For neutrons, the scattering 
length is unnaturally large with $a_{nn} = - 18.5 \pm 0.3 \fm$ (for a recent 
review see~\cite{Phillips}).
The effective range is expected to be approximately charge-independent,
and thus $r_{nn} = 2.7 \fm$~\cite{effrange}. Consequently, the universal
regime is restricted to extremely low neutron densities $n = \kf^3/
(3\pi^2) < 10^{-4} \fm^{-3}$.

Dilute Fermi gases with resonant interactions were realized for 
the first time by O'Hara {\it et al.}~in 2002~\cite{Thomas1}. In this
and subsequent experiments, the universal factor $\xi$ was determined 
by extracting an equation of state from the properties of the atomic 
cloud~\cite{Thomas1,Thomas2,Salomon,Grimm}. This
leads to $\xi = 0.51 \pm 0.04$~\cite{Thomas2}, 
$\xi \approx 0.7$~\cite{Salomon} and 
$\xi = 0.27\,^{+0.12}_{-0.09}$~\cite{Grimm}, for temperatures in units
of the Fermi temperature $T/T_{\rm F} \approx 0.05$ (except for 
$T/T_{\rm F} \approx 0.6$ in~\cite{Salomon}). To date, the most reliable 
theoretical results for the universal equation of state are from $T=0$, 
fixed-node Green's function Monte Carlo simulations, $\xi = 0.44 \pm 
0.01$~\cite{Carlson} and $\xi = 0.42 \pm 0.01$~\cite{Astra}.

The virial expansion has been applied to cold atoms in the vicinity 
of Feshbach resonances by Ho {\it et al.}~\cite{Ho1,Ho2}.
The resulting universal equation of state for high temperatures
describes many observed properties of resonant Fermi gases, with 
a universal virial coefficient that is independent
of temperature. In this paper, we study to what extent the virial
equation of state of low-density neutron matter approaches this 
unitary limit.

There are many theoretical approaches to low-density neutron matter.  
Recently, the equation of state of neutron matter at zero temperature 
has been calculated using an effective field theory for large scattering 
length and large effective range~\cite{dEFT}. This provides a 
model-independent description of neutron matter for densities where 
the interparticle separation is comparable to the effective range. 
The effective field theory calculation is particularly transparent,
since the couplings are fitted directly to the scattering length and the 
effective range. 

In addition, there are microscopic calculations starting 
from nucleon-nucleon (NN) and three-nucleon interactions 
that reproduce NN scattering and selected few-nucleon data. One of 
these are Hartree-Fock calculations~\cite{RGnm} with the model-independent 
low-momentum interaction $\vlk$~\cite{Vlowk}. These results are
promising, because low-momentum interactions offer the possibility of a
perturbative and thus systematic approach to nucleonic matter~\cite{nucmat}.
Moreover, for neutron matter at subnuclear densities, the Hartree-Fock 
results agree with complicated Fermi hyper-netted chain~\cite{FP,APR} and 
Brueckner~\cite{Brueckner1,Brueckner2} calculations. However, these 
approaches may not be reliable for very low densities, where 
the large scattering length physics is important. This is
especially problematic for Brueckner calculations, since the hole-line 
expansion relies on Pauli blocking effects being dominant, so that all 
large structures are dissolved in the medium. Note that the equation of
state calculation of Buchler and Coon~\cite{BC} is close in strategy to 
the virial equation of state, but it takes into account Pauli blocking
on the phase shifts and thus uses a model NN interaction.
There are also a number of Skyrme-type~\cite{skyrme1,skyrme2,skyrme3} 
and relativistic mean-field~\cite{RMF1,RMF2,RMF3} 
parametrizations of the nuclear energy functional,
which are used to calculate ground-state energies and densities of
intermediate-mass and heavy nuclei. However, the resulting energy 
functional for neutron matter is not well constrained from fits to
finite nuclei.

Finally, there are promising lattice simulations for neutron matter 
using effective field theory, where the couplings are regularized on 
the same lattice~\cite{Dean}. These calculations are presently limited 
to small lattices and to the lowest-order contact interaction, which is 
fitted to reproduce the scattering length. In this paper, we study
how the virial equation of state depends on NN scattering properties 
and in particular on the physics beyond the large scattering length. 
As a result, we assess how a low-order truncation in the effective field 
theory impacts the equation of state of low-density neutron matter. 
This provides a valuable check for the lattice results.

The virial equation of state is a general, model-independent equation of state 
for a dilute gas, provided the fugacity $z = e^{\mu/T}$ is small. Here $\mu$ 
denotes the chemical potential. An additional assumption is that the 
system is in a gas phase and has undergone no phase transition with 
decreasing temperature or increasing density. Under these conditions,
the partition function can be expanded in powers of the fugacity.
The second virial coefficient $b_2$ describes the $z^2$ term in this
expansion and is directly related to the two-body scattering phase 
shifts~\cite{b2,huang}. Little is known about the third virial 
coefficient, which describes the $z^3$ term~\cite{b3Pais,b3Bedaque}. 
We emphasize that the virial expansion is not a perturbative
expansion in powers of $\kf \as$. A great advantage of the virial 
formalism is that it includes both bound states and scattering resonances 
on an equal footing. It correctly predicts that thermodynamic
quantities, such as the pressure, are continuous as the interaction is
changed to convert a low-energy scattering resonance into a weakly-bound
state. This continuity has been demonstrated experimentally with cold atoms 
in the crossover region of a Feshbach resonance (see e.g.,~\cite{Salomon}) 
and was also shown theoretically by Ho and Mueller using the viral
expansion across $\as = \pm \infty$~\cite{Ho1}.

This paper is organized as follows. We briefly introduce the virial 
equation of state in Section~\ref{veos}. Further details can be found
in~\cite{vEOSnuc}. In Section~\ref{results}, we present results for
the second virial coefficient, the pressure, energy, entropy and the free
energy of low-density neutron matter. 
We also study the dependence of the virial 
coefficient on NN scattering properties and show that neutron matter 
scales to a very good approximation. Finally, we conclude in 
Section~\ref{conclusions}.

\section{Virial Equation of State}
\label{veos}

For the virial equation of state we expand the grand-canonical partition
function or the pressure in a power series of the fugacity
\be
P=\frac{2T}{\lambda^3}\bigl( z + z^2 b_n + z^3 b_n^{(3)} + \mathcal{O}(z^4) 
\bigr) \,,
\label{p}
\ee
where $\lambda$ denotes the nucleon thermal wavelength $\lambda = 
(2\pi/mT)^{1/2}$ and $b_n$, $b_n^{(3)}$ are the second and third 
virial coefficients for neutron matter respectively. We will include
$b_n^{(3)}$ only to make an error estimate for our virial results.
The density follows from differentiating the pressure with respect to 
the fugacity $n = z/T \, (\partial_z P)_{V,T}$ and is given by
\be
n = \frac{2}{\lambda^3}\bigl( z + 2 z^2 b_n + 3 z^3 b_n^{(3)} + 
\mathcal{O}(z^4) \bigr) \,.
\label{n}
\ee
Therefore, the fugacity expansion is an expansion in powers
of $n \lambda^3$, and for finite temperatures, it is valid to much higher
densities than the $\kf \as$ expansion.
In this work, we 
truncate the virial expansion after second order in the fugacity. This
leads to an equation of state that is thermodynamically consistent.

The dependence of the density on $z$ can be inverted. This gives the 
virial equation of state directly in terms of density and temperature
$P = P\bigl(z(n,T),T\bigr)$. In practice, we directly, without this
inversion, generate the equation of state in tabular form for a range 
of fugacity values. This maintains the thermodynamic consistency of the 
virial equation of state.

The second virial coefficient is related to the partition function of 
the two-particle system $\sum_{\text{states}} e^{-E_{2}/T}$, where the 
sum is over all two-particle states of energy $E_{2}$.  This sum can be 
converted to an integral over relative momentum $k$ weighted by the 
density of states of the interacting two-particle system. The density
of states, and therefore the second virial coefficient, can then be 
expressed in terms of the scattering phase shift $\delta(k)$, summed over 
all allowed partial waves~\cite{b2,huang}.

For neutron matter, the second virial coefficient $b_n$ is given by
\begin{equation}
b_n(T)=\frac{1}{2^{1/2}\,\pi T} \int_0^\infty dE \: e^{-E/2T} \, 
\delta^{\text{tot}}(E) - 2^{-5/2} \,,
\label{bn}
\end{equation}
where $-2^{-5/2}$ is the free Fermi gas contribution and $\delta^{
\text{tot}}(E)$ is the sum of the isospin-triplet elastic scattering
phase shifts at laboratory energy $E$. This sum is over all partial waves
with two-particle spin $S$, angular momentum $L$ and total angular
momentum $J$ allowed by spin statistics, and includes a degeneracy 
factor $(2J+1)$,
\begin{align}
\delta^{\text{tot}}(E) &= \sum_{S,L,J} (2J+1) \, 
\delta_{\,^{2S+1}\text{L}_J}(E)
\nonumber \\[1mm]
&=\delta_{^1\text{S}_0}+\delta_{^3\text{P}_0}+3\,\delta_{^3\text{P}_1}
+5\,\delta_{^3\text{P}_2}+5\,\delta_{^1\text{D}_2}+ \ldots
\label{deltantot}
\end{align}
Note that we have neglected the effects of the mixing parameters due to 
the tensor force. We expect that their contributions to the second virial 
coefficient describing spin-averaged observables vanish.

Finally, the entropy $S$ and the energy $E$ are obtained from the virial 
equation of state using thermodynamics~\cite{Ho1,huang}. 
The entropy density $s=S/V$ follows from differentiating the pressure
with respect to the temperature $s=(\partial_T P)_{\mu}$. This leads to
\be
s = \frac{5P}{2T} - n \log z + \frac{2T}{\lambda^3} z^2 b_n^\prime \,,
\label{s}
\ee
where $b_n^\prime(T)=db_n(T)/dT$ denotes the temperature derivative of the 
virial coefficient. The energy density $\epsilon=E/V$ can be calculated from
the entropy density,
\be
\epsilon=T s + n \mu - P \nonumber
=\frac{3}{2}P - \frac{2T^2}{\lambda^3} z^2 b_n^\prime \,.
\label{epsilon}
\ee
For completeness, the entropy per particle $S/N$, energy per particle 
$E/N$ and the free energy per particle $F/N$ are given by
\begin{equation}
\frac{S}{N} = \frac{s}{n} \: , \quad
\frac{E}{N} = \frac{\epsilon}{n} \quad \text{and} \quad
\frac{F}{N} = \frac{f}{n} \,,
\label{esfa}
\end{equation}
with the free energy density $f=\epsilon - T s$.

\section{Results}
\label{results}

\subsection{Virial Coefficients}

\begin{figure}[t]
\begin{center}
\includegraphics[scale=0.36,clip=]{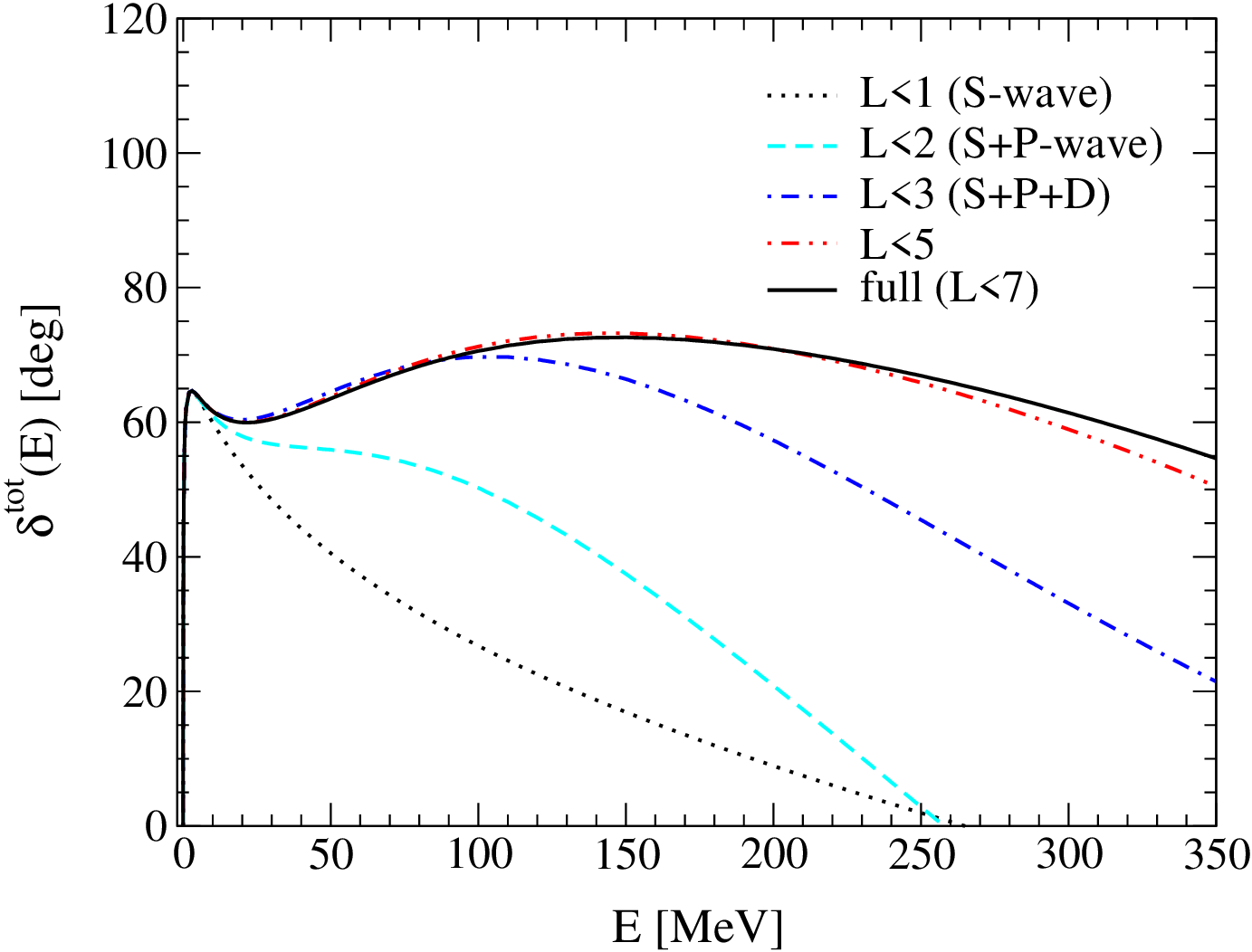}
\caption{(Color online) The total phase shift $\delta^{\text{tot}}(E)$ 
versus laboratory energy $E$ obtained from including successively higher 
partial waves.}
\label{Fig1}
\end{center}
\end{figure}

We first calculate the second virial coefficient $b_n$. We use the 
$T=1$ $np$ 
phase shifts obtained
from the Nijmegen partial wave analysis~\cite{nnphases}. This
neglects the small charge dependences in nuclear interactions. We have
included all partial waves with $L \leqslant 6$ and the resulting total
total phase shift $\delta^{\text{tot}}(E)$ is shown in Fig.~\ref{Fig1}.
The total phase shift displays both the prominent low-energy $^1$S$_0$ 
resonance and the significance of higher partial waves. We observe that 
$\delta^{\text{tot}}(E)$ is very weakly dependent on energy 
with $\delta^{\text{tot}} \approx 65$ degrees over the entire range
$E \leqslant 350 \mev$ of the Nijmegen partial wave analysis.
As shown in Fig.~\ref{Fig1},
the decrease of the $^1$S$_0$ phase shift due to the large effective 
range is compensated by the contributions from higher angular momenta.
For reference, in the unitary limit of a S-wave resonance ($a_s = \pm 
\infty$), the phase shift is energy-independent with $\delta_{^1\text{S}_0} 
= 90$ degrees.

\begin{table*}
\caption{The second virial coefficient $b_n$ for different temperatures. 
In addition to the full results, we also give results for the virial 
coefficient calculated in the unitary limit and from including only 
the S-wave scattering length $a_{np}$, with and without the effective 
range $\re$ contribution 
($a_{np} = -23.768 \fm$ and $\re = 2.68 \fm$~\cite{effrange}).
The results labeled CIB take into account the effects due to
charge-independence breaking (CIB) on the scattering length with 
$a_{nn} = -18.5 \fm$.}
\label{Table1}
\begin{ruledtabular}
\begin{tabular}{c|ccccccc}
$T [\text{MeV}]$ & $\as = \pm \infty$ & $\as = \pm \infty, \re$ &
$a_{np}$ & $a_{np}, \re$ & $b_n$ full & $b_n$ full with CIB & 
$T \, b_n^\prime$ full \\[0.2mm] \hline
 1 & 0.530 & 0.449 & 0.357 & 0.287 & 0.288 & 0.251 & 0.032 \\[0.2mm]
 2 & 0.530 & 0.417 & 0.400 & 0.298 & 0.303 & 0.273 & 0.012 \\[0.2mm]
 3 & 0.530 & 0.394 & 0.421 & 0.296 & 0.306 & 0.279 & 0.004 \\[0.2mm]
 4 & 0.530 & 0.376 & 0.434 & 0.291 & 0.306 & 0.283 & 0.001 \\[0.2mm]
 5 & 0.530 & 0.360 & 0.443 & 0.285 & 0.306 & 0.285 & 0.000 \\[0.2mm]
 6 & 0.530 & 0.347 & 0.450 & 0.278 & 0.306 & 0.286 & 0.001 \\[0.2mm]
 7 & 0.530 & 0.335 & 0.456 & 0.272 & 0.307 & 0.288 & 0.002 \\[0.2mm]
 8 & 0.530 & 0.324 & 0.460 & 0.265 & 0.307 & 0.289 & 0.004 \\[0.2mm]
 9 & 0.530 & 0.314 & 0.464 & 0.259 & 0.308 & 0.291 & 0.007 \\[0.2mm]
10 & 0.530 & 0.305 & 0.467 & 0.254 & 0.309 & 0.292 & 0.009 \\[0.2mm]
12 & 0.530 & 0.289 & 0.472 & 0.243 & 0.310 & 0.295 & 0.013 \\[0.2mm]
14 & 0.530 & 0.275 & 0.476 & 0.233 & 0.313 & 0.299 & 0.017 \\[0.2mm]
16 & 0.530 & 0.263 & 0.479 & 0.224 & 0.315 & 0.302 & 0.020 \\[0.2mm]
18 & 0.530 & 0.252 & 0.482 & 0.216 & 0.318 & 0.305 & 0.022 \\[0.2mm]
20 & 0.530 & 0.242 & 0.484 & 0.208 & 0.320 & 0.308 & 0.023 \\[0.2mm]
22 & 0.530 & 0.233 & 0.486 & 0.201 & 0.322 & 0.311 & 0.023 \\[0.2mm]
24 & 0.530 & 0.224 & 0.488 & 0.194 & 0.324 & 0.313 & 0.022 \\
\end{tabular} 
\end{ruledtabular}
\end{table*}

In the unitary limit, the second virial coefficient is independent of
the temperature and given by $b_2 = 3/2^{5/2} = 0.53$~\cite{Ho1}. For neutron 
matter, our results for the virial coefficient $b_n$ and its derivative
$T b_n^\prime(T)$ are listed in Table~\ref{Table1}. In addition to the 
full results, we also give the virial coefficient obtained only from
the S-wave scattering length $a_{np}$ and including the effective range
contribution. As expected, $b_n$ is dominated by the large S-wave 
scattering length, but effective range and higher partial wave 
contributions are significant even for these low temperatures.
As a result of the weak energy dependence of $\delta^{\text{tot}}(E)$,
we find that the second virial coefficient is approximately independent 
of temperature over a wide range, and consequently 
$T b_n^\prime \approx 0$. The value we obtain for 
$b_n \approx 0.31$ is $40 \%$ reduced compared to the universal value 
$b_2 = 0.53$. 

In Table~\ref{Table1}, we also 
study the effects of charge-independence breaking (CIB) 
on the scattering length. Due to the lack of neutron-neutron scattering
data, we estimate CIB effects by subtracting the virial coefficient
calculated only with $a_{np}$ from the full $b_n$, and then add the 
virial coefficient obtained from the neutron-neutron scattering 
length $a_{nn} = -18.5 \fm$. We find that CIB effects are largest for 
$T < 5 \mev$ and lead to a $10 \%$ reduction of the second virial 
coefficient.

Finally, we extend our results for $b_n$ to temperatures $T \geqslant
25 \mev$ in Table~\ref{Table2}. We give these results separately, because 
there is a small error for these higher temperatures due to the truncation of 
the integration over the phase shifts at $E \leqslant 350 \mev$ (the
extent of the partial wave analysis). Assuming
the total phase shift is constant, we vary the energy
cutoff to $E > 350 \mev$ and estimate the error to be $< 5 \%$ 
up to the highest temperatures given in Table~\ref{Table2}.

\begin{table}[t]
\caption{The second virial coefficient $b_n$ for temperatures $T \geqslant
25 \mev$. As discussed in the text, we estimate an error of $< 5 \%$ 
for these higher temperatures due to the truncation of the integration 
over the phase shifts at $E \leqslant 350 \mev$.}
\label{Table2}
\begin{ruledtabular}
\begin{tabular}{c|cc}
$T [\text{MeV}]$ & $b_n$ full & $T \, b_n^\prime$ full \\[0.2mm] \hline
25 & 0.325 & 0.022 \\[0.2mm]
30 & 0.329 & 0.015 \\[0.2mm]
35 & 0.330 & 0.004 \\[0.2mm]
40 & 0.330 & -0.009 \\[0.2mm]
45 & 0.328 & -0.025 \\[0.2mm]
50 & 0.324 & -0.041 \\
\end{tabular} 
\end{ruledtabular}
\end{table}

\subsection{Pressure}

\begin{figure}[t]
\begin{center}
\includegraphics[scale=0.36,clip=]{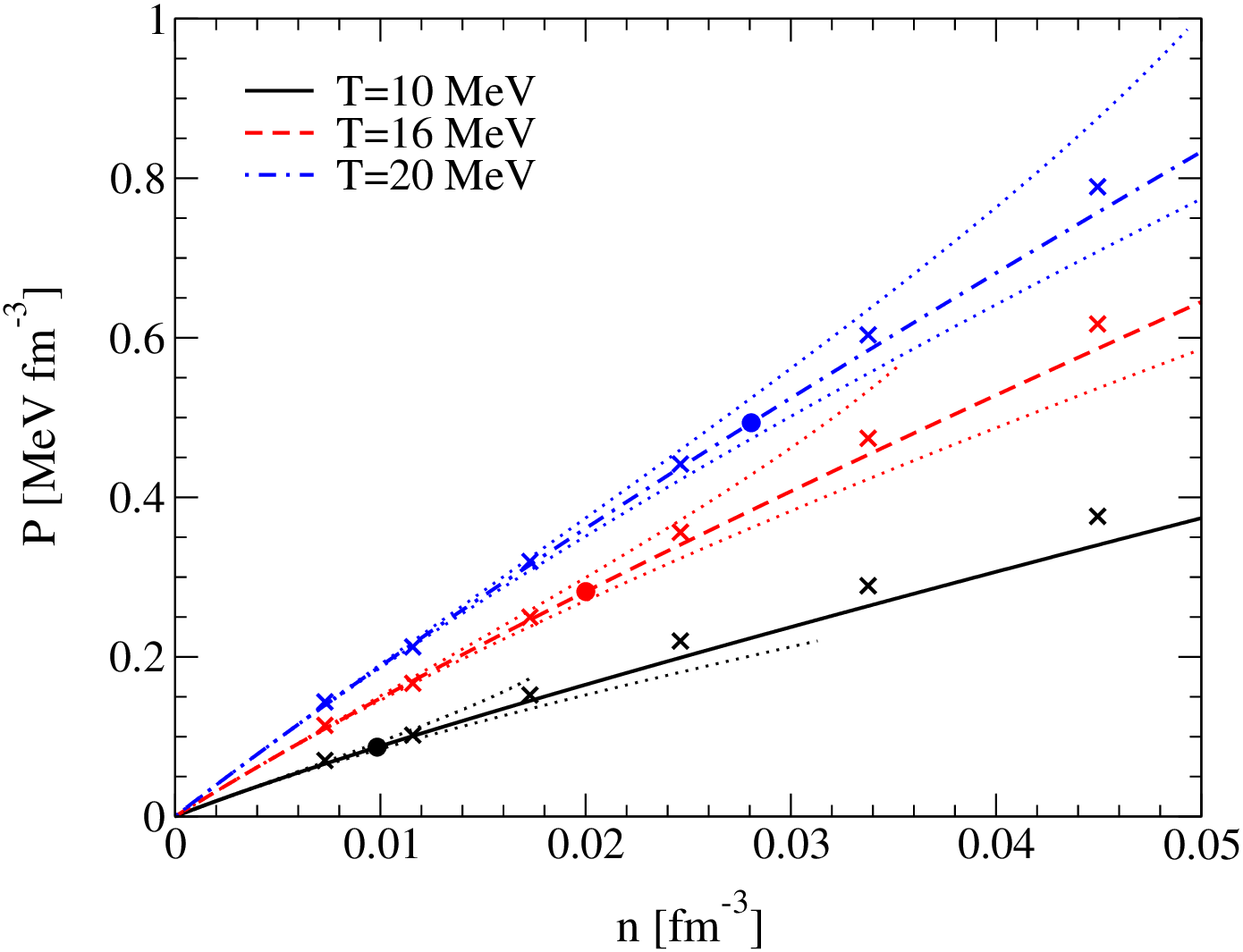}
\caption{(Color online) The pressure $P$ versus density $n$ 
for $T=10$, $16$ and $20 \mev$. The dotted error bands for the 
virial equation of state are based on an estimate of a neglected 
third virial coefficient $b^{(3)}_n = \pm b_n/2$. Note that a 
negative $b_n^{(3)}$ increases the pressure. Also shown are 
the FHNC results of Friedman and Pandharipande (crosses)~\cite{FP}. 
The circles indicate where the fugacity is $z =0.5$, and the error
bands are shown for $z \leqslant 1$. For this density range, the
fugacities are $z<1.65$, $1.01$ and $0.79$ for $T=10$, $16$ and 
$20 \mev$.}
\label{Fig2}
\end{center}
\end{figure}

Our virial results for the pressure are shown in Fig.~\ref{Fig2} for
temperatures $T=10$, $16$ and $20 \mev$. The third virial coefficient 
can be used to make a simple error estimate of neglected terms in the 
virial expansion. The Pauli principle prevents three neutrons to interact
in the S-wave and there is no three-neutron bound state. Moreover, 3N
forces are very small in low-density neutron matter. Therefore,
we consider a third virial coefficient $|b_n^{(3)}| 
\lesssim b_n/2$ reasonable. 
Note that for the ideal Fermi gas $b_n^{(3)} = 3^{-5/2} = 0.06$. 
In Fig.~\ref{Fig2}, we show the resulting error bands, which are small 
even up to $z \leqslant 1$. We also compare our results to the
the microscopic Fermi hyper-netted-chain (FHNC) equation of state of 
Friedman and Pandharipande~\cite{FP}. We find that the FHNC results
are in very good agreement with the model-independent virial equation 
of state. Moreover, it is intriguing that the FHNC results lie within
the error band for the virial pressure even for fugacities $z > 0.5$.
In contrast, we have found a disagreement of the FHNC results with 
the virial equation of state for low-density nuclear matter due to 
clustering~\cite{vEOSnuc}.

\subsection{Scaling and Energy}

\begin{figure}[t]
\begin{center}
\includegraphics[scale=0.36,clip=]{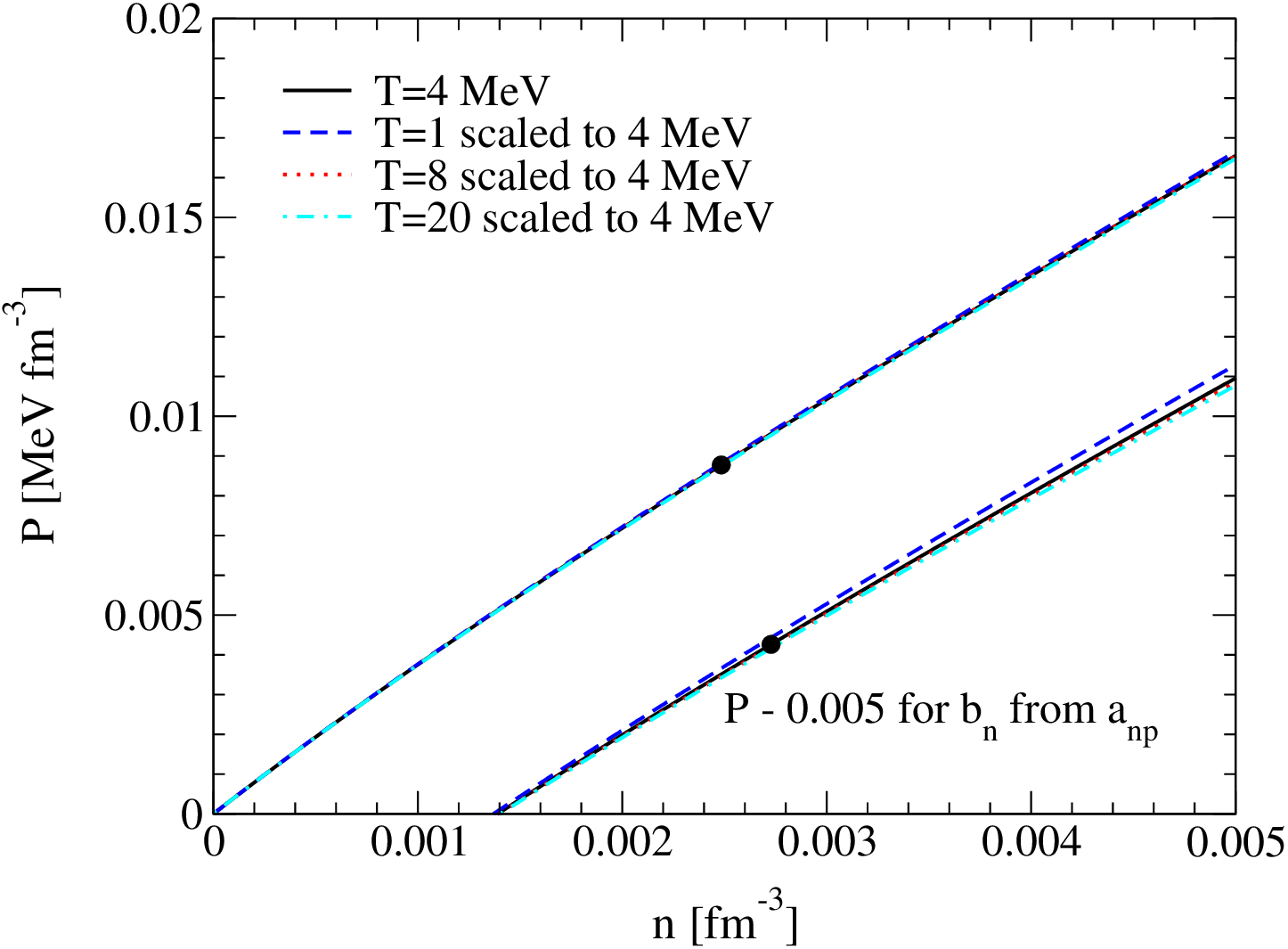}
\caption{(Color online) The pressure $P$ versus density $n$
calculated for $T=1$, $4$, $8$ and $20 \mev$ and scaled to a 
temperature of $T'=4 \mev$ (for details see text). The lower
curves are results for the scaled pressures obtained with a 
virial coefficient calculated only from the S-wave scattering length
$a_{np}$. The latter curves have been shifted for better readability.
The circles indicate where the fugacity is $z =0.5$, and for this
density range, the fugacity is $z<0.87$.}
\label{Fig3}
\end{center}
\end{figure}

\begin{figure}[t]
\begin{center}
\includegraphics[scale=0.36,clip=]{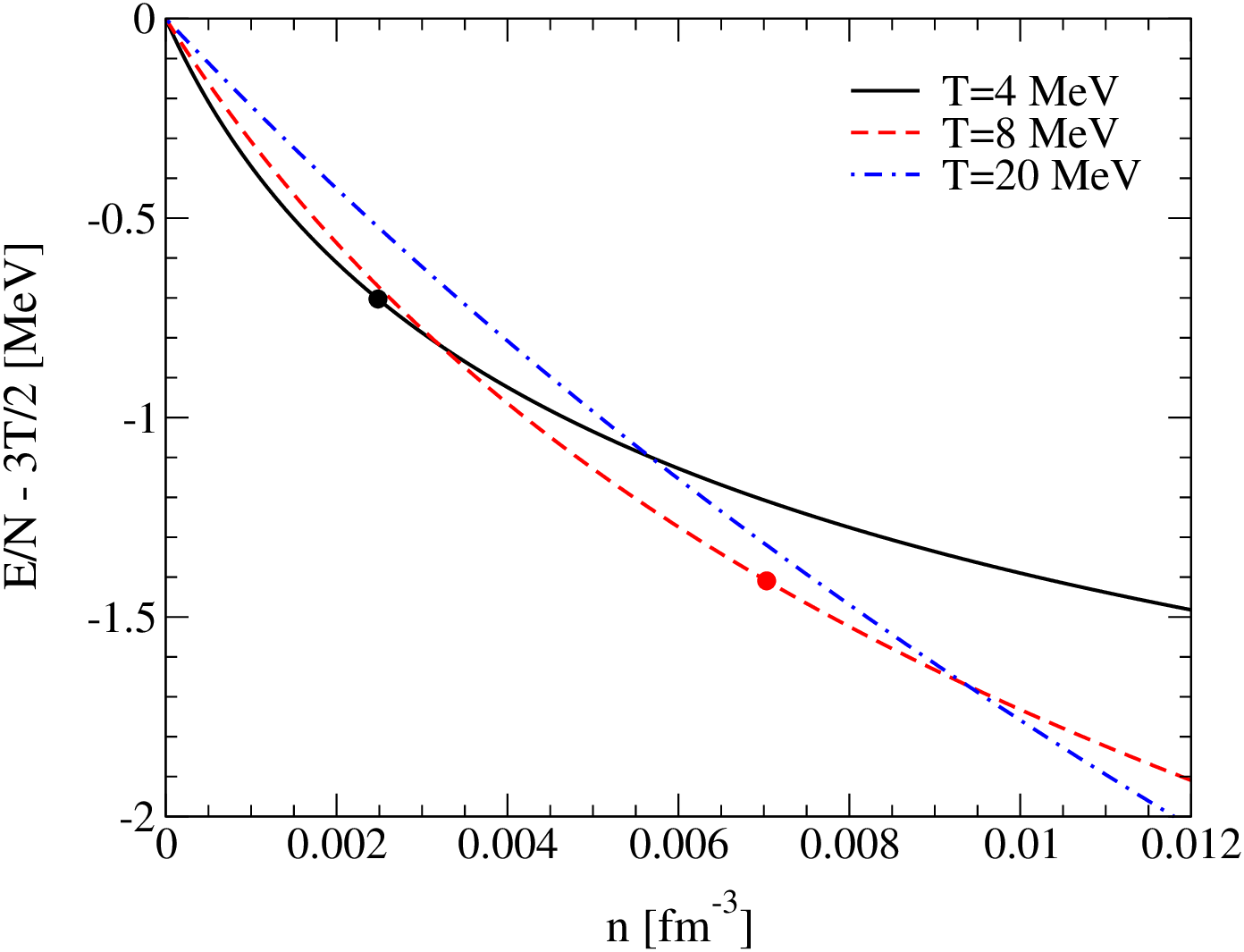}
\caption{(Color online) The energy per particle $E/N$ versus
density $n$ for $T=4$, $8$ and $20 \mev$. In order to clearly 
show the interaction effects we have subtracted the free kinetic 
energy $3T/2$ from all curves. The energy per particle
obtained using the universal relation $\epsilon = 3P/2$ with
the corresponding virial pressures are indistinguishable from
these curves. The circles indicate where the fugacity is $z =0.5$,
and for this density range, the fugacities are $z<1.60$, 
$0.77$ and $0.25$ for $T=4$, $8$ and $20 \mev$.}
\label{Fig4}
\end{center}
\end{figure}

If the virial coefficients are temperature independent, then the power series 
in the fugacity for the pressure and the density will have no explicit 
temperature dependence. Consequently, the dependence of the pressure on
density and temperature will be given by
\be
P(n,T) = T^{5/2} \, f(n/T^{3/2}) \,,
\ee
where $f(x)$ is a scaling function of $n/T^{3/2}$. If this scaling relation 
holds, one can predict the pressure $P(n',T')$ at a new temperature $T'$ 
from $P(n,T)$ through
\be
P(n',T')=\biggl(\frac{T'}{T}\biggr)^{5/2} P(n,T) \,,
\label{pscale}
\ee
with $n'=(T'/T)^{3/2} \, n$. In Fig.~\ref{Fig3}, we demonstrate that 
low-density neutron matter scales according to Eq.~(\ref{pscale}).
The predicted pressures for $T'=4 \mev$ obtained from virial pressures 
calculated for $T=1$, $8$ and $20 \mev$ are in excellent agreement with 
the unscaled $T=4 \mev$ virial pressure. Fig.~\ref{Fig3} also gives
results for the scaled pressures ($T=1$, $8$ and $20 \mev$ scaled to 
$T'=4 \mev$ and unscaled $T=4 \mev$) obtained from the virial equation of 
state with $b_n$ calculated only from the S-wave scattering length $a_{np}$.
Although Table~\ref{Table1} shows that $b_n$ increases by $35 \%$ from
$T=1$ to $20 \mev$ in this case, the scaling symmetry continues 
to hold to a good approximation.

In the scaling regime, $b_n^\prime = 0$ and therefore the energy
density, Eq.~(\ref{epsilon}), is given by
\be
\epsilon = 3 P/2 \,.
\label{uniepsilon}
\ee
This is a general thermodynamic relation in the universal 
regime~\cite{Ho}. In Fig.~\ref{Fig4}, we show the energy per 
particle for $T=4$, $8$ and $20 \mev$. In order to separate
the interaction effects we have subtracted the free kinetic 
energy $3T/2$ from all curves. We have also calculated the
energy per particle from the virial pressures using the
universal relation Eq.~(\ref{uniepsilon}). The resulting
curves for the energy per particle are indistinguishable from
those shown in Fig.~\ref{Fig4}. This
demonstrates that the $b_n^\prime$ term is indeed negligible.
While neutron matter scales as in the unitary limit, the
thermodynamic properties depend on the value of the virial
coefficient, and the latter depends on the physics of NN 
scattering.

Finally, we emphasize that the universal relation for the energy
density, Eq.~(\ref{uniepsilon}), holds independent of the value
of the universal interaction coefficient given by $\xi$ for $T=0$
and $b_n$ for higher temperatures. In fact, Eq.~(\ref{uniepsilon})
is valid for $T=0$, where the energy per particle only scales with
density, $E/N \sim \xi \, n^{2/3}$ according to Eq.~(\ref{unieos}), and
for the virial regime, where the energy per particle scales with
density and temperature $E/N \sim \xi(T/n^{2/3}) \, n^{2/3}
\sim \widetilde{\xi}(T/n^{2/3}) \, T$, with Fermi temperature
$T_{\rm F} \sim n^{2/3}$. The latter scaling is very
explicit to lowest order in the 
density, where the energy per particle is given by
\be
\frac{E}{N} \approx \frac{3}{2} \, T \, ( 1 - b_n \lambda^3 n/2 ) \,.
\ee

\subsection{Entropy and Free Energy}

\begin{figure}[t]
\begin{center}
\includegraphics[scale=0.36,clip=]{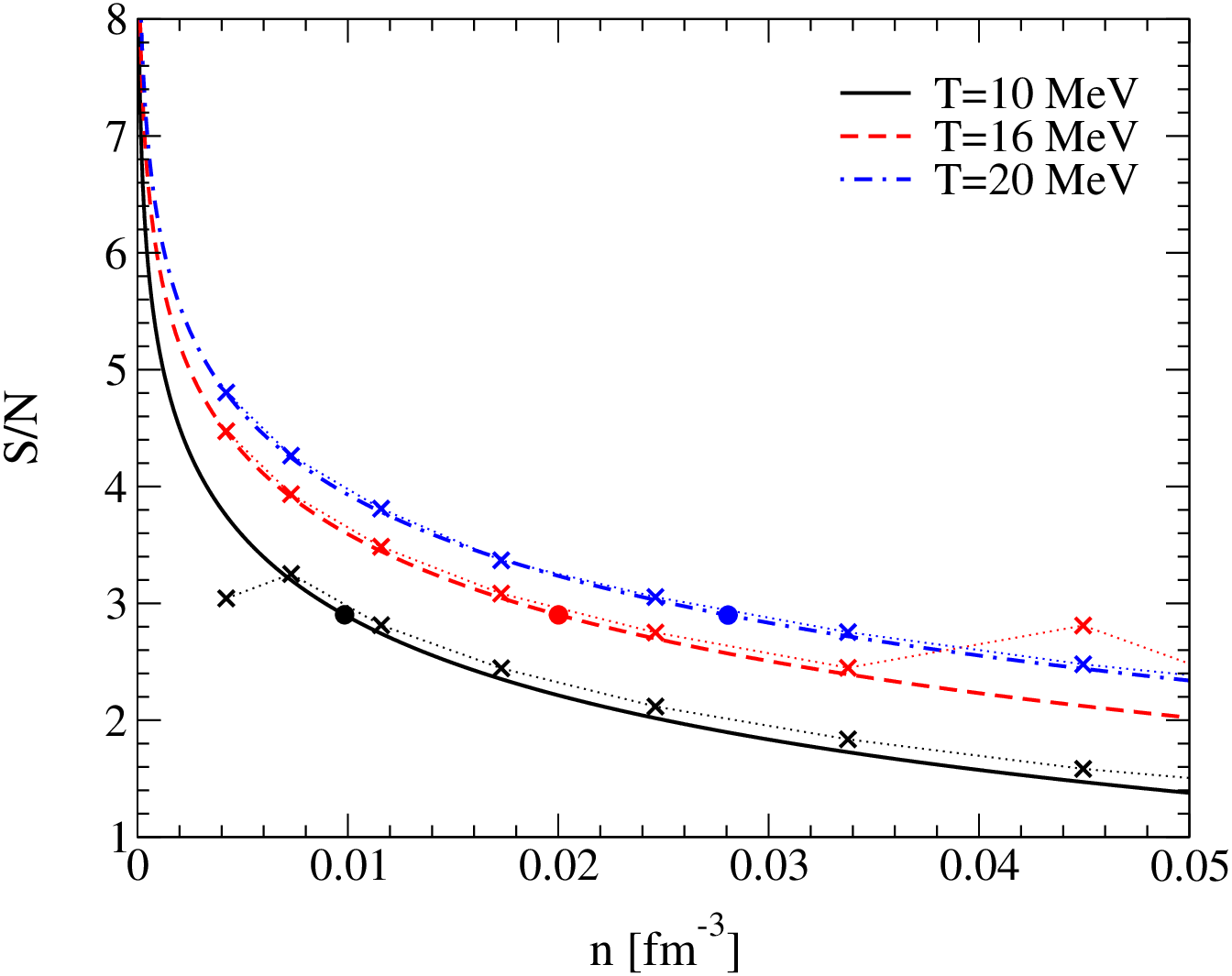}
\caption{(Color online) The entropy per particle $S/N$ 
versus density $n$ for $T=10$, $16$ and $20 \mev$.
Also shown are the FHNC results of Friedman and Pandharipande
(crosses)~\cite{FP}. The circles indicate where the fugacity is $z =0.5$.}
\label{Fig5}
\end{center}
\end{figure}

\begin{figure}[t]
\begin{center}
\includegraphics[scale=0.36,clip=]{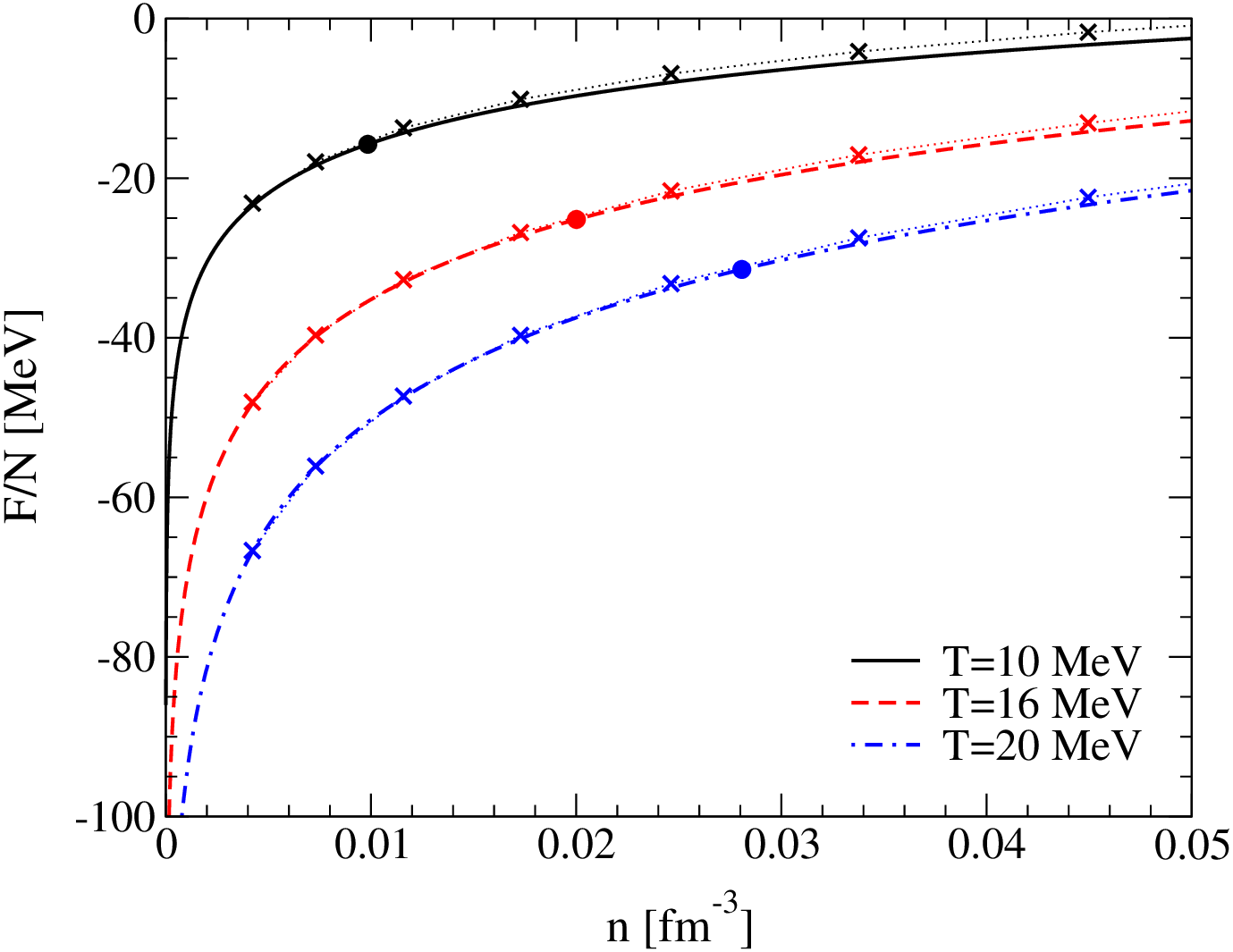}
\caption{(Color online) The free energy per particle $F/N$ 
versus density $n$ for $T=10$, $16$ and $20 \mev$.
Also shown are the FHNC results of Friedman and Pandharipande
(crosses)~\cite{FP}. The circles indicate where the fugacity is $z =0.5$.}
\label{Fig6}
\end{center}
\end{figure}

We present our virial results for the entropy per particle in
Fig.~\ref{Fig5} for temperatures $T=10$, $16$ and $20 \mev$. 
The FHNC results are again in very good agreement with
the model-independent virial equation of state, with the exception
of two FHNC points that seem to either reflect numerical instabilities or
are typos in~\cite{FP}. As indicated by the circles for $z=0.5$ in 
Fig.~\ref{Fig5}, we observe that the range of validity of the virial
equation of state is bounded by a constant entropy per particle,
independent of the temperature. This is easily understood from the 
expression for the entropy per particle in the scaling regime 
with $b_n^\prime = 0$,
\be
\frac{S}{N} = \frac{5}{2} \, 
\biggl(1- \frac{z b_n}{1+2z b_n}\biggr) - \log z \,.
\ee
Therefore, if one takes as the range of validity of the virial equation 
of state $z \lesssim 0.5$, this is equivalent to $S/N \gtrsim 2.9$ for
$b_n = 0.31$, in agreement with Fig.~\ref{Fig5}. For completeness, we also
show the free energy per particle in Fig.~\ref{Fig6}. As for the pressure
and the entropy, the FHNC results agree well with our model-independent
virial predictions.

\section{Conclusions}
\label{conclusions}

We have presented the virial equation of state of low-density neutron
matter. The virial expansion provides a systematic way to include
strong interactions with a well-defined range of validity. The resulting
virial equation of state has a simple parametric form and is thermodynamically 
consistent. We have calculated the second virial coefficient directly 
from the NN scattering phase shifts. Therefore, the virial equation of 
state is model-independent and sets a benchmark for all nuclear equations 
of state at low densities.

We have found that the second virial coefficient $b_n$ is approximately 
temperature-independent. This is the result of an intriguing cancellation.
The decrease of the $^1$S$_0$ phase shift due to the effective range is 
compensated by the increase of the higher angular momentum phase shifts. 
Therefore, while $b_n$ is generally dominated by the low-energy $^1$S$_0$ 
resonance, contributions from higher partial waves are significant even
for these low temperatures. The virial coefficient $b_n \approx 0.31$ is
reduced by $40 \%$ compared to the unitary limit, where $b_2 = 0.53$.

The virial equation of state was used to make model-independent predictions 
for the pressure, energy, entropy and the free energy of low-density neutron 
matter over a wide range of densities and temperatures. Our results include 
the physics of the large neutron scattering length in a tractable way.
The range of validity of the virial equation of state is given by the
fugacity. With $z \lesssim 0.5$, one has for the entropy per particle
$S/N \gtrsim 2.9$ or for the density $n \lesssim 4 \cdot 10^{11} \, 
(T/\text{MeV})^{3/2} \gcq$. The virial expansion thus provides important 
constraints on the physics of the neutrinosphere in supernovae. 

We have made simple error estimates of the virial equation of state by 
studying the effects of a neglected third virial coefficient $|b_n^{(3)}|
\lesssim b_n/2$. The resulting error bands are small. For a better error 
estimate, it is important to have a reliable calculation of the third 
virial coefficient. The FHNC results of Friedman and Pandharipande~\cite{FP} 
are in very good agreement with the model-independent virial equation of 
state. This is in contrast to our findings for low-density nuclear 
matter~\cite{vEOSnuc}. The FHNC and virial results agree within the
estimated errors even for densities, where $z > 0.5$.

The temperature independence of $b_n$ leads to a scaling symmetry of
low-density neutron matter. As in the universal regime, thermodynamic 
properties of neutron matter scale over a wide range of temperatures.
This extends the approximate scaling of the $T=0$ equation of 
state according to Eq.~(\ref{coldneut}) to finite temperatures. 
Finally, model-independent predictions for the density and spin response 
of low-density neutron matter and a detailed comparison with nuclear 
lattice calculations will be discussed in future work.

\section*{Acknowledgments}

This work is supported by the US Department of Energy
under Grant No. DE--FG02--87ER40365 and the National Science Foundation
under Grant No. PHY--0244822.

\end{document}